\documentclass{article}
\usepackage{spconf,amsmath,graphicx}
\usepackage{multirow}
\usepackage{booktabs}

\usepackage{float}
\usepackage[dvipsnames]{xcolor}
\definecolor{blue}{rgb}{0.0, 0.5, 0.8}

\title{On granularity of prosodic representations in expressive text-to-speech}
\name{Mikolaj Babianski, Kamil Pokora, Raahil Shah, Rafal Sienkiewicz, Daniel Korzekwa, Viacheslav Klimkov}
\address{Amazon Text-to-Speech Research \\ \small\texttt{\{babiansk, kamipoko, vklimkov\}@amazon.com}}
\copyrightnotice{978-1-6654-7189-3/22/\$31.00~\copyright2023 IEEE}
\begin{document}
%
\maketitle
\begin{abstract}
In expressive speech synthesis it is widely adopted to use latent prosody representations to deal with variability of the data during training.
Same text may correspond to various acoustic realizations, which is known as a one-to-many mapping problem in text-to-speech.
Utterance, word, or phoneme-level representations are extracted from target signal in an auto-encoding setup, to complement phonetic input and simplify that mapping.
 This paper compares prosodic embeddings at different levels of granularity and examines their prediction from text. We show that utterance-level embeddings have insufficient capacity and phoneme-level tend to introduce instabilities when predicted from text. Word-level representations impose balance between capacity and predictability. As a result, we close the gap in naturalness by 90\% between synthetic speech and recordings on LibriTTS dataset, without sacrificing intelligibility.
\end{abstract}
\section{Introduction}
\label{sec:introduction}
Neural Text-to-Speech (NTTS) \cite{shen2018natural} is characterized by synthesizing speech waveform solely with deep neural networks. 
This paradigm greatly enhanced naturalness and flexibility of speech synthesis.
It enables new applications such as expressive \cite{prateek2019other,ezzerg2021enhancing} and low-resource \cite{shah21_ssw} speech generation, speaker identity \cite{karlapati2020copycat} and prosody transplantation \cite{klimkov2019fine, Bilinski2022_interspeech}.
This paper focuses on expressive speech synthesis, i.e. generation of speech that originally contains great degree of variation in terms of intonation and inflections.

This variation is not described by phoneme sequence, typically used as input to NTTS. Thus, the statistical model has to perform a one-to-many mapping, where the same input text can correspond to different acoustic realizations. Vanilla modelling approaches suffer from averaging and fail to reproduce the original variability of the training data.

To avoid averaging, it is common to use additional input that describes variability in the data. Initially, it was proposed to extract a single latent representation of the target speech in an auto-encoder manner for the whole utterance \cite{skerry2018towards}. Target speech is not available during inference, so either the centroid representation is used \cite{zhang2019learning} or it is separately predicted from text \cite{stanton2018predicting,karlapati2021prosodic}. A single representation for the whole utterance can't store temporal information effectively, thus, it was proposed to use more fine-grained representations at the phoneme-level for the task of prosody transplantation \cite{klimkov2019fine,lee2019robust}. This idea was further expanded to text-to-speech, where word-level \cite{hodari2021camp,klapsas2021word} and phoneme-level \cite{ren2020fastspeech,elias2021parallel,liu2021delightfultts} representations were utilized. At the fine-grained level, prosody can be represented with pre-extracted features such as pitch, energy, spectral tilt, but learnt representations can convey more information and represent more abstract aspects of prosody such as emotions. Therefore, in the rest of the paper we focus on learnt representations.

This paper provides a systematic comparison of prosodic representations at different levels of granularity. We compare performance of utterance, word, and phoneme-level prosody embeddings in terms of \emph{a)} capacity: what if we have a perfect prosody predictor; \emph{b)} predictability: how sensitive is the approach to inaccurate prosody predictions.
Main contributions of this study are:
\begin{itemize}
    \item We systematically compare prosody embeddings at different levels of granularity.
    \item A solution to intelligibility issues in the case of phoneme-level prosody reference is proposed.
    \item We show the trade-off between capacity and predictability of prosody embeddings, advocating the use of word-level representations.
    \item We examine data quantity and input features needed for robust prosody prediction from text.
\end{itemize}

The rest of the paper is organized as follows: Section \ref{sec:acoustic_model} describes the text-to-speech framework used; Section \ref{sec:prosody_embeddings_predictor} elaborates on prosody embedding prediction from text; Section \ref{sec:experiments_prosody_embeddings_granularity} compares prosody embeddings at different levels of granularity in objective and subjective evaluations; Section \ref{sec:experiments_prosody_prediction} presents ablation studies on prosody embedding prediction; Section \ref{sec:conclusions} concludes the paper.

\begin{figure*}[htb]
\vspace{-3mm}
\begin{minipage}[b]{.48\linewidth}
  \centering
  \centerline{\includegraphics[width=8.0cm]{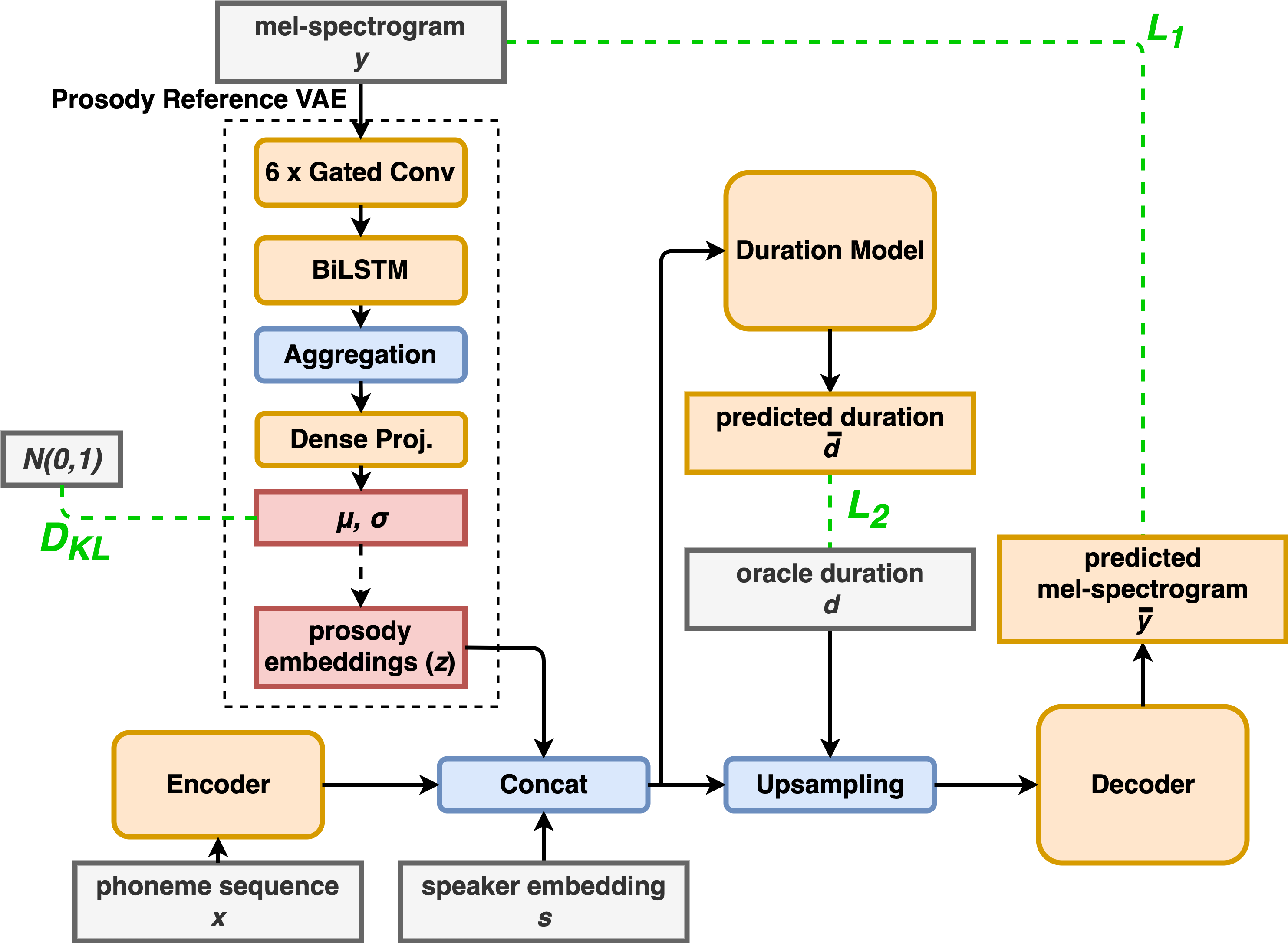}}
  \centerline{\textit{(a)}}\medskip
\end{minipage}
\hfill
\begin{minipage}[b]{0.48\linewidth}
  \centering
  \centerline{\includegraphics[width=8.0cm]{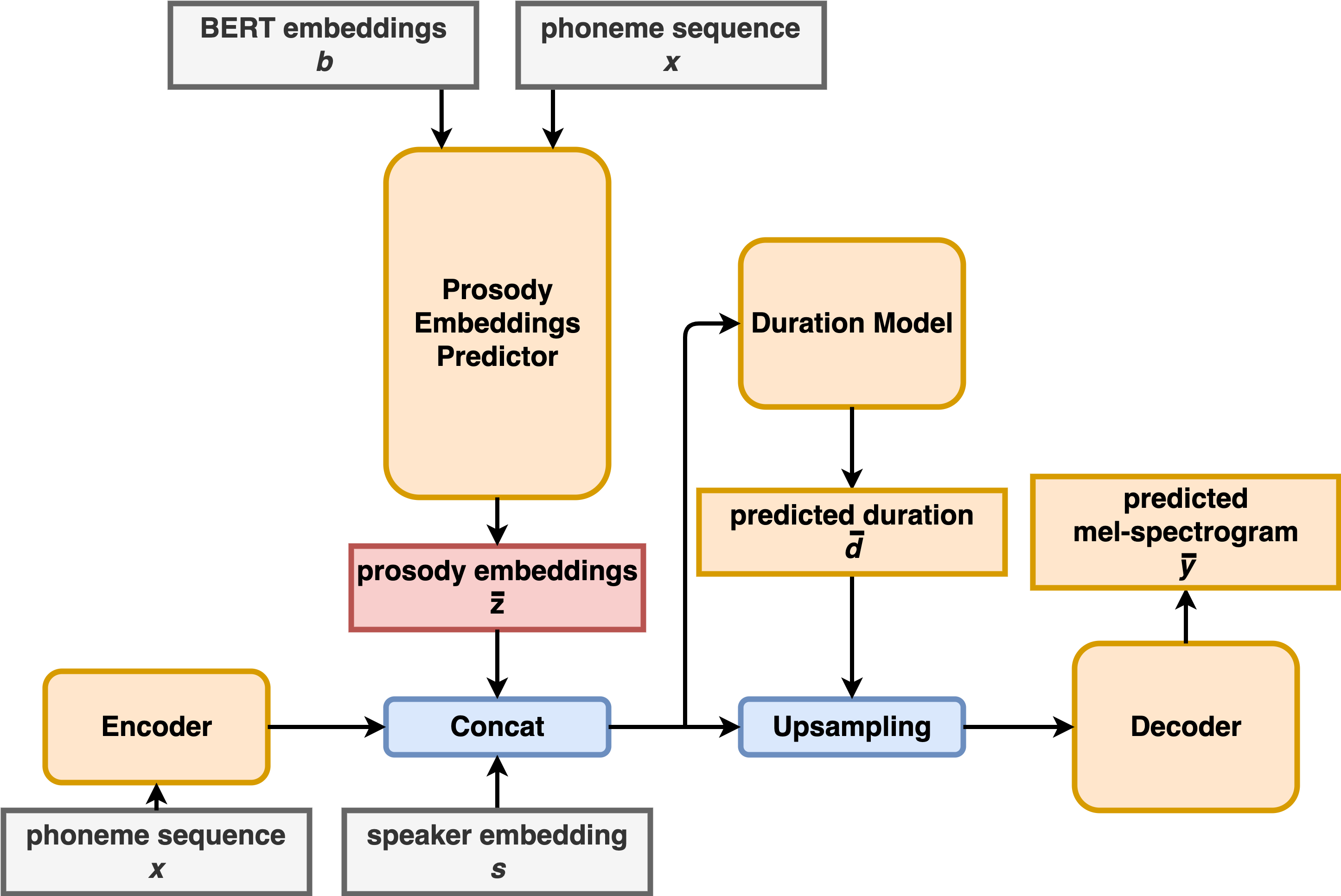}}
  \centerline{\textit{(b)}}\medskip
\end{minipage}
\vspace{-5mm}
\caption{\textit{Schematic diagram of the TTS model during a) training and b) inference. The dashed arrow denotes sampling from parametric distribution. Components in red are of prosody embeddings granularity (utterance/word/phoneme). Green, dashed lines denote loss functions.}}
\vspace{-3mm}
\label{fig:acoustic_model}
\end{figure*}

\section{Acoustic Model}
\label{sec:acoustic_model}
The backbone of our acoustic model architecture (Figure \ref{fig:acoustic_model}) is similar to the explicit duration TTS model presented in Shah et al. \cite{shah21_ssw}.
It follows the encoder-decoder paradigm, where the input phoneme sequence \(\boldsymbol{x}\) is encoded by a phoneme encoder presented in the Tacotron2 \cite{shen2018natural} paper. We concatenate the encoded phoneme sequence with both speaker \(\boldsymbol{s}\) and prosody \(\boldsymbol{z}\) embeddings upsampled by repetition \cite{ren2020fastspeech,elias2021parallel} to the phoneme-level.
Speaker embeddings are represented as corresponding entries in the embedding look-up table. Prosody embeddings are obtained via compression of the mel-spectrogram \(\boldsymbol{y}\) with the use of variational prosody reference encoder described in Section \ref{ssec:acoustic_model_reference_encoder}. 
During inference, the encoded sequence is upsampled accordingly to alignments produced by the duration model, described in Section \ref{ssec:acoustic_model_duration_model}. The upsampled sequence is then passed to the decoder to map the disentangled linguistic features, speaker and prosodic contents into acoustic parameters represented as mel-spectrograms. In this work, we use the non-autoregressive decoder presented in Shah et al. \cite{shah21_ssw}.

\subsection{Variational Prosody Reference Encoder}
\label{ssec:acoustic_model_reference_encoder}
To alleviate the one-to-many problem of TTS we use the variational prosody reference encoder \cite{elias2021parallel}. We aim to learn the latent representation of the information, which cannot be derived from the other input streams - phoneme sequence and speaker embedding. For clarity of the architecture presentation, here we describe only one level of granularity - word-level. Modification of the model architecture to adjust for different prosody embedding granularities is described in Section \ref{ssec:acoustic_model_prosody_embeddings_granularity}.
The variational reference encoder (Figure \ref{fig:acoustic_model}a) takes target mel-spectrogram frames as input and converts them into a sequence of $n$ latent vectors \(\boldsymbol{z}\), which corresponds to the number of words in the utterance. We refer to this representation as word-level prosody embeddings. 

The encoder comprises a stack of six residual gated convolution blocks \cite{oord2016wavenet}. Each residual gated convolution block is composed of a 1D-convolution with a kernel size of 15 and a hidden dimension of 512, followed by a $tanh$ filter and a $sigmoid$ activation gate which are element-wise multiplied and then added to a residual connection. The convolution stack is followed by a BiLSTM layer with a hidden dimension of 128.
We use a dropout of 0.1 in convolutional and BiLSTM layers. The BiLSTM layer output is firstly aggregated to the word-level by taking a middle frame of each word.
Then, after a dense projection we obtain a sequence of Gaussian distribution parameters \(\boldsymbol{\mu}\) and \(\boldsymbol{\sigma}\), which we use to sample a sequence of prosody embeddings corresponding to words \(\boldsymbol{z}\) of dimension 8. Finally, we upsample the word-level prosody embeddings by repetition to the phoneme-level and concatenate them with the phoneme encoder output (Figure \ref{fig:acoustic_model}a).

As we do not have access to target mel-spectrograms at inference time (Figure \ref{fig:acoustic_model}b), a separate model is introduced to predict prosody embeddings \(\boldsymbol{z}\) from text. The architecture for this model is described in Section \ref{sec:prosody_embeddings_predictor}.

\subsection{Duration Model}
\label{ssec:acoustic_model_duration_model}
Neural TTS requires learning the alignment between two different length sequences, which are the text, represented by phonemes, and speech, represented by acoustic parameters i.e. mel-spectrogram frames. There are two major approaches to obtain the alignment: attention-based \cite{shen2018natural} and explicit-duration-based \cite{shah21_ssw, ren2021fastspeech,elias2021parallel,shen2020non}.
Attention-based components typically used for this task have known instabilities, which exhibit in synthesised speech as mumbling, early cut-offs, word repetition and word skipping \cite{he2019robust,guo-interspeech-2019,battenberg-icassp-2020}.
Following recent research in the field inspired by traditional parametric speech synthesis techniques \cite{heiga-spss,heiga-spss-classic}, these issues are mitigated by explicitly modelling the durations of phonemes \cite{shah21_ssw, ren2021fastspeech,elias2021parallel,shen2020non} which we decide to use in this work.

We use forced alignment from a Gaussian Mixture Model (GMM) based external aligner in the Kaldi Speech Recognition Toolkit \cite{Povey_ASRU2011} to produce ground truth duration for each phoneme, represented as the integer number of mel-spectrogram frames it corresponds to.
To predict these durations, we train a duration model component following the architecture detailed in Shah et al. \cite{shah21_ssw}, with the addition of speaker and prosody embeddings conditioning.
The duration model is trained jointly with the acoustic model by minimizing L2 loss function in the logarithmic domain between predicted and ground truth phoneme durations.
During training, teacher forcing is used, i.e. the acoustic model uses ground truth duration values to upsample a phoneme's encoding to the respective number of mel-spectrogram frames.

\subsection{Prosody Embeddings Granularity}
\label{ssec:acoustic_model_prosody_embeddings_granularity}
The model described above uses word-level prosody embeddings. Which means that there is one embedding corresponding to each word in the input text.
In this work we also explore two other levels of prosody modelling granularity: phoneme-level (one embedding per each phoneme) and utterance-level (single embedding for the whole utterance).
For the phoneme-level prosody modelling we change the reference encoder to output one embedding per each phoneme and reduce prosody embedding dimension from 8 to 3, which we found optimal in terms of stability, for this level of granularity.
In the case of utterance-level prosody modelling we use a stride of 2 in the residual gated convolution blocks of the reference encoder in order to gradually downsample the time resolution \cite{wang2018style}.
Then, we project the first and last state of the BiLSTM layer into two vectors of dimension 64, which represent mean \(\boldsymbol{\mu}\) and standard deviation \(\boldsymbol{\sigma}\) of the posterior distribution.
Finally we sample a single 64-dimensional prosody embedding \(\boldsymbol{z}\) corresponding to the whole utterance.

\subsection{Training Procedure}
\label{ssec:acoustic_model_training_procedure}
To train the acoustic and duration models we use Adam optimiser with $\beta_1 = 0.9$  and $\beta_2 = 0.98$. We use a linear warm-up of the learning rate from 0.1 to 1 for the first 10k steps, followed by an exponential decay from 10k steps to 100k steps with a minimum value of $10^{-5}$.
Acoustic and duration models are trained jointly for 500K steps with a batch-size equal 32 and are optimized with respect to the following loss function:
\begin{equation}
L_{total} = L_{1 melspectrogram} + L_{2 logduration} + \gamma * D_{KL}
\end{equation}
where $L_{1 melspectrogram}$ is the $L_{1}$-distance between predicted and oracle mel-spectrograms and $L_{2 logduration}$ is the $L_{2}$-distance between predicted and ground truth durations calculated in the logarithmic domain.  $D_{KL}$ is the Kullback–Leibler divergence between outputs of the variational prosody reference encoder and $N(0,1)$. We find the optimal value of $\gamma$ to be $10^{-3}$ for the phoneme-level and $10^{-5}$ for the utterance and word-level prosody modelling. We present ablation of the $\gamma$ parameter in Section \ref{ssec:experiments_stability}.

\vspace{-1mm}
\section{Prosody Embeddings Predictor}
\begin{figure}[H]                                                         
\hspace*{-5mm}                                     
\centering                                                             
\includegraphics[width=80mm]{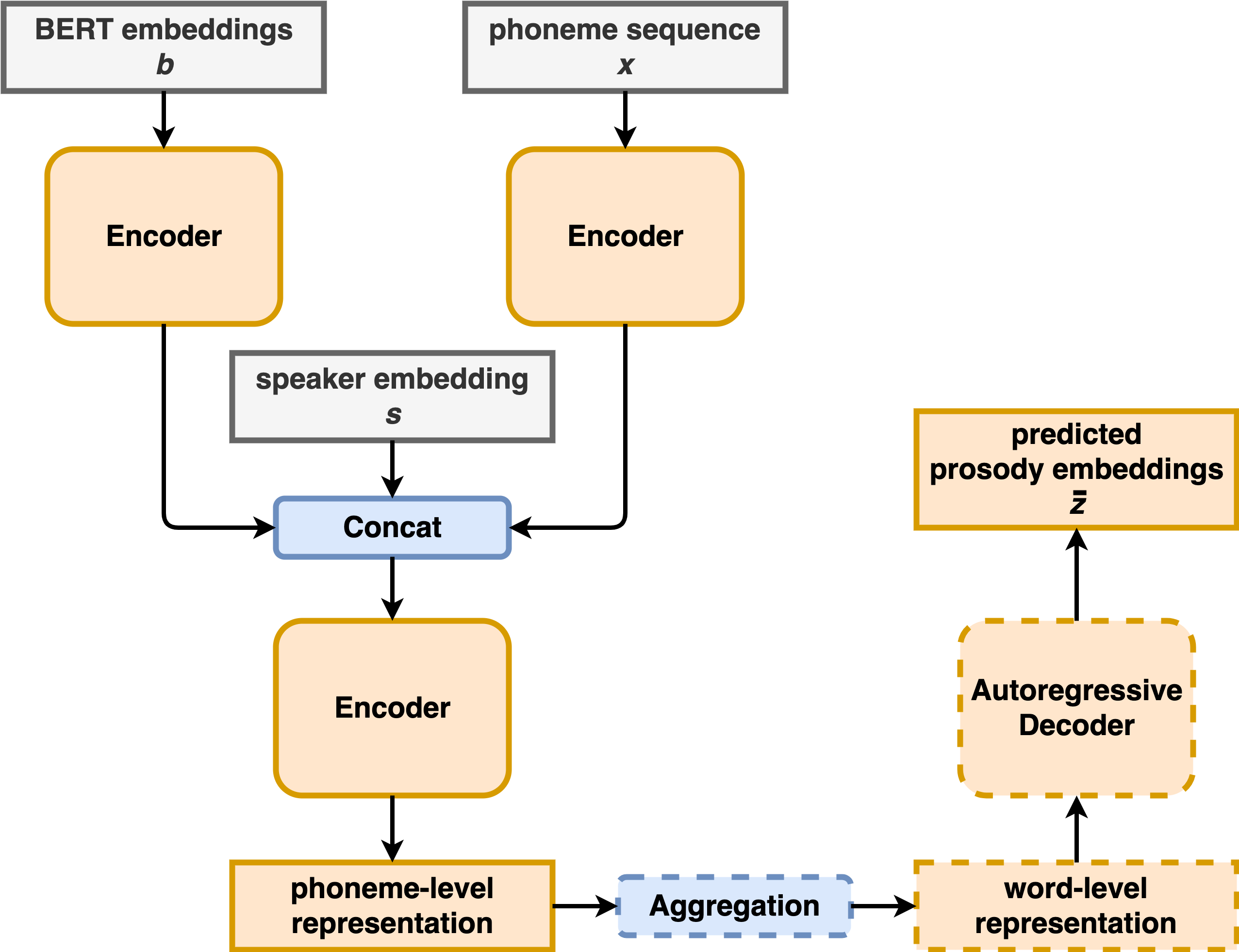}
\vspace{-2mm}
\caption {\textit{Schematic diagram of the prosody embeddings predictor model. Components with dashed border are used only for fine-grained (word or phoneme-level) prosody embeddings prediction. } }
\label{fig:prosody_predictor}
\end{figure}

\vspace{-1mm}
\label{sec:prosody_embeddings_predictor}
At inference time we do not have access to target mel-spectrograms, therefore, we use a separate model to predict prosodic representations \(\boldsymbol{z}\) from text (Figure \ref{fig:acoustic_model}b).
The prosody embeddings predictor model (Figure \ref{fig:prosody_predictor}) has three input streams: phoneme sequence, contextual word embeddings extracted with a pre-trained \(\rm{BERT}\) model \cite{devlin2018bert} and speaker embedding.
Phoneme sequence and contextual word embeddings are encoded by separate Tacotron2-based encoders.
We upsample the encoded \(\rm{BERT}\) embeddings from the word-level to the phoneme-level and the speaker embedding to the phoneme-level, before concatenating them with the encoded phoneme representations. Next, we pass the concatenated sequence to another Tactotron2-based encoder block.
After that, in the case of word-level embeddings prediction, the encoded phoneme-level representation is aggregated to the word-level by taking the middle frame of each word.
Finally, we use an autoregressive decoder to predict prosody embeddings.
The autoregressive decoder is inspired by the architecture of the Tacotron2 mel-spectrogram decoder.
In order to adapt the decoder to the prosody prediction task, we reduce the hidden dimension of the LSTM-layers to 128 and pre-net hidden dimension to 6. To predict the utterance-level prosody representation, instead of the autoregressive decoder, we use a simple linear projection layer.

We train this model using prosody embeddings extracted with previously trained acoustic model as target labels.
Specifically, the model is trained to predict posterior mean \(\boldsymbol{\mu}\) for each target prosody embedding using L2 loss and teacher-forcing framework.

\vspace{-3mm}

\section{Experiments - Prosody Embeddings Granularity}
\label{sec:experiments_prosody_embeddings_granularity}
In this section we conduct a systematic study of prosodic representations at different levels of granularity applied to the expressive TTS task.
The performance of utterance, word, and phoneme-level prosody embeddings is compared in terms of capacity and predictability.
We evaluate naturalness as well as stability and intelligibility of synthesized speech.

\subsection{Data}
\label{ssec:experiments_data}
Evaluations are conducted on a publicly available corpus of audiobook recordings - LibriTTS \cite{zen2019libritts}.
From which we use only recordings marked as clean.
The training set consists of approximately 250 hours of speech (split into 140,000 utterances) narrated in an expressive manner by 1229 speakers.
For validation we use a held-out set of 1000 randomly selected utterances from the 100 most frequent speakers.
We extract 80-band mel-spectrograms with a 12.5 ms frame-shift as acoustic features.

\subsection{Systems}
\label{ssec:experiments_systems}
We use our acoustic model (Section \ref{sec:acoustic_model}) along with the prosody embeddings predictor (Section \ref{sec:prosody_embeddings_predictor}) to test 3 levels of prosody embeddings granularity:
1) \textit{G-VAE} - utterance-level.
2) \textit{W-VAE} - word-level.
3) \textit{P-VAE} - phoneme-level.
All mel-spectrogram prediction systems are used in combination with the Universal Neural Parallel WaveNet Vocoder \cite{jiao2021universal} in order to obtain a 24kHz audio signal.

\subsection{Subjective Evaluation Protocol}
\label{ssec:experiments_subjective_evaluation_protocol}
For the subjective evaluation we conduct MUSHRA tests \cite{itu20031534} with the Amazon Mechanical Turk platform.
60 native English speakers are presented with the samples in a random order side-by-side, and are asked to \textit{``Evaluate naturalness of the samples on the scale from 0 to 100.''} 
A total of 1000 utterances are used for testing and the test is balanced in such a way that each test case is scored by 3 listeners independently. 
Ground truth mel-spectrograms vocoded with the Universal Parallel WaveNet Vocoder (\textit{Ref} system) are used as an upper anchor. 
The significance of the MUSHRA results is analyzed using a Wilcoxon signed-rank test with Bonferroni-Holm correction applied \cite{clark2007statistical}.

\subsection{Stability}
\label{ssec:experiments_stability}
To quantify the stability of tested systems and the intelligibility of synthesized speech we conduct Word Error Rate (WER) analysis.
The whole test set of 1000 utterances described in Section \ref{ssec:experiments_data} is used for the evaluation.
We transcribe speech generated in the TTS mode (prosody embeddings predicted from text) with the ASpIRE Chain ASR model from Kaldi.
Then the WER is computed between the sentence text and the corresponding transcription.

The G-VAE and W-VAE models have WER scores (Table \ref{table:wer}) comparable to recordings when trained with \(D_{KL}\) loss weight ($\gamma$) equal to \(10^{-5}\).
Whereas, training the P-VAE model in an analogical setup results in significant stability issues.
We believe that this is caused by the phoneme-level prosody embeddings distribution being very hard to predict from text.
Only after applying more regularization during training by increasing \(D_{KL}\) loss weight we are able to obtain a phoneme-level model matching other systems in terms of WER score.
Intelligibility is a crucial property of TTS system.
Therefore, for all the other experiments we use the G-VAE and W-VAE models trained with \(\gamma=10^{-5}\) and the P-VAE model trained with \(\gamma=10^{-3}\), as they are matching our stability requirements. 
We found that further increasing $\gamma$ parameter does not bring any significant improvements and may lead to degradation in segmental quality of synthesized audio.

\begin{table}[htb]
\centering
\begin{tabular}{lll}
\toprule
\textbf{System}                 & \boldmath{$D_{KL} \gamma$} & \textbf{WER $\downarrow$}      \\ \hline
\textbf{G-VAE}                  & $1e-5$                   & $2.18\% \pm 0.30$ \\ \hline
\textbf{W-VAE}                  & $1e-5$                   & $2.13\% \pm 0.30$ \\ \hline
\multirow{4}{*}{\textbf{P-VAE}} & $1e-5$                   & $3.59\% \pm 0.36$ \\ \cline{2-3} 
                                & $1e-4$                   & $2.47\% \pm 0.31$ \\ \cline{2-3} 
                                & $1e-3$                   & $2.17\% \pm 0.29$ \\ \cline{2-3} 
                                & $1e-2$                   & $2.17\% \pm 0.29$ \\ \hline
\textbf{Ref}                    & $-$                      & $2.29\% \pm 0.31$ \\ 
\bottomrule
\end{tabular}
\caption{\textit{Word Error Rate with 95\% confidence intervals \cite{bisani2004bootstrap} computed across the 1000 test utterances, along with \(D_{KL}\) loss weight (\(\gamma\)).}}
\label{table:wer}
\end{table}

\begin{figure*}[t]
\begin{minipage}[b]{.48\linewidth}
  \centering
  \centerline{\includegraphics[width=7.0cm]{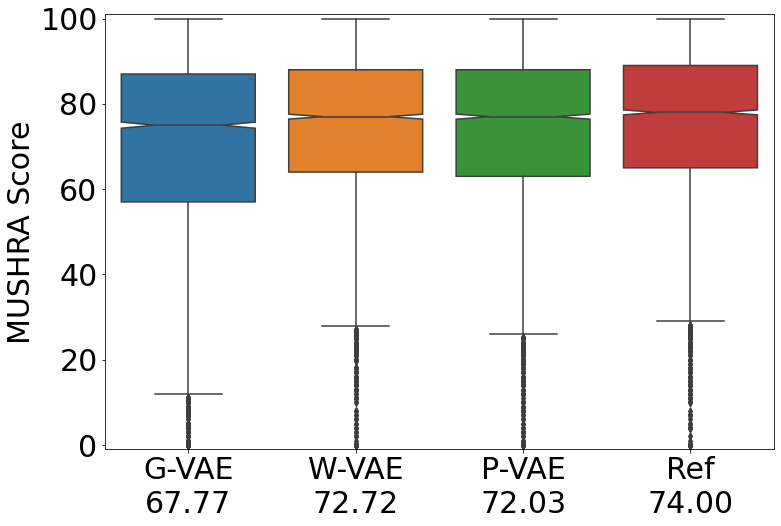}}
  \centerline{\textit{(a) Ground truth prosody embeddings}}\medskip
\end{minipage}
\hfill
\begin{minipage}[b]{0.48\linewidth}
  \centering
  \centerline{\includegraphics[width=7.0cm]{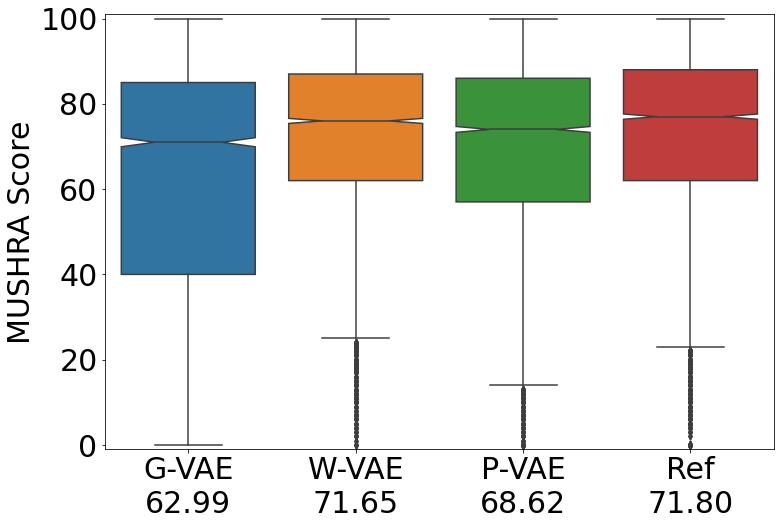}}
  \centerline{\textit{(b) Prosody embeddings predicted from text}}\medskip
\end{minipage}
\caption{\textit{Subjective listeners ratings from the naturalness MUSHRA tests with a) ground truth prosody embeddings and b) prosody embeddings predicted from text (TTS). Mean scores are reported below the system names.}}
\label{fig:mushra_naturalness}
\end{figure*}

\subsection{Capacity}
\label{ssec:experiments_oracle_resynthesis}
We analyse the best-case performance of acoustic models by simulating perfectly predicted prosody embeddings in the Oracle Resynthesis setup. 
That is, at inference time we provide ground truth latent representations from the variational reference encoder for all tested systems.
We evaluate naturalness in a subjective test as described in Section \ref{ssec:experiments_subjective_evaluation_protocol} and summarize results in Figure \ref{fig:mushra_naturalness}a.
In this setup, the G-VAE model scores significantly lower than the other systems ($p$-$value<0.01$), suggesting that fine-grained embeddings are required for natural prosody modelling.
There is no statistically significant difference between W-VAE and P-VAE systems ($p$-$value>0.01$).
It is worth mentioning, that we have also conducted an analogical experiment with a P-VAE model trained with lower \(D_{KL}\) loss weight ($\gamma=10^{-5}$).
With such model, when ground truth prosody embeddings provided during inference, we are able to reconstruct speech almost perfectly.
However, it comes at a cost of stability issues, when prosody embeddings predicted from text are used as described in Section \ref{ssec:experiments_stability}.

To gain a deeper understanding of prosody representations, we evaluate our acoustic models in the Oracle Resynthesis setup with changed speaker embedding.
That is, we extract prosody from a source recording and resynthesize the same text with changed voice - a so-called Voice Conversion (VC).
We convert all 1000 test utterances into 4 selected (two male and two female) target voices.

To effectively measure how close the prosody patterns of converted speech are to the source recordings, we first extract fundamental frequency (F0) at the frame-level from source and converted audio pairs with the RAPT algorithm \cite{talkin1995robust}. Then we calculate two metrics commonly used to measure the linear dependence of prosody contours: Pearson Correlation Coefficient (PCC) and Root Mean Squared Error (RMSE) \cite{sisman2020overview}. We can see that the results (Table \ref{table:vc}) are very similar for the W-VAE and P-VAE models, while the G-VAE performs much worse in both metrics. This reinforces the subjective evaluation findings that utterance-level embeddings do not provide sufficient capacity to capture expressive prosody and fine-grained modelling is required for expressive speech generation.

\begin{table}[htb]
\centering
\begin{tabular}{lll}
\toprule
\textbf{System}          & \textbf{F0 RMSE $\downarrow$} & \textbf{F0 PCC $\uparrow$} \\ \hline
\textbf{G-VAE}           & $1.693 \pm 0.012$                  & $0.760 \pm 0.003$                  \\ \hline
\textbf{W-VAE}           & $1.535 \pm 0.011$                  & $0.801 \pm 0.002$                  \\ \hline
\textbf{P-VAE}           & $1.539 \pm 0.012$                  & $0.802 \pm 0.003$                  \\
\bottomrule 
\end{tabular}
\caption{\textit{Objective Voice Conversion evaluation metrics with 95\% confidence intervals computed between source and converted speech: F0 Root Mean Square Error (RMSE), F0 Pearson Correlation Coefficient (PCC) \cite{sisman2020overview}.}}
\label{table:vc}
\end{table}

\subsection{Predictability}
\label{ssec:experiments_text_to_speech_naturalness}
Finally, we evaluate our systems in the TTS scenario.
That is, we generate speech from textual input only and provide prosody embeddings predicted from text during inference.
The naturalness of synthesized speech is evaluated subjectively as described in Section \ref{ssec:experiments_subjective_evaluation_protocol}.
The results are summarised in Figure \ref{fig:mushra_naturalness}b (all comparisons are statistically significant with $p$-$value<0.01$).
The G-VAE model scores much worse than other systems.
It fails to reproduce expressive speech variability and tends to output flat prosody contours due to averaging.
We also observe issues with accurate phoneme duration prediction for the G-VAE model, e.g. unnatural, fast-paced speech.
We hypothesise, that a single representation for the whole utterance can’t store temporal information effectively.
Such issues are not visible in the models using fine-grained prosody representations.
While both of them score much higher than the G-VAE, the word-level model performs better in terms of fidelity and prosody naturalness and closes the gap between the P-VAE model and the reference system by over 90\%.
We conclude that word-level representations impose a good compromise for expressive prosody modelling granularity.
They have enough capacity to produce varied and natural speech and still can be accurately predicted from text.

\section{Experiments - Prosody Prediction}
\label{sec:experiments_prosody_prediction}
In this section we focus on semantically concerted prosody prediction. 
First, we investigate the impact of data quantity used in the training procedure. 
Second, we conduct an ablation study of the prosody embeddings predictor input streams.

\subsection{Data Quantity}
\label{ssec:experiments_data_quantity}
We investigate the impact of training data quantity on our system, by looking into a single-speaker scenario with limited amount of data.
We build a dataset by taking 15000 utterances from the HiFi corpus \cite{bakhturina2021hi} coming from one male speaker (id 6097).
Again, we keep a held-out set of 1000 randomly selected utterances for validation.
We train the word-level acoustic model and the prosody embeddings predictor in two setups: 
1) HiFi - using only 15000 utterances from a single HiFi speaker.
2) HiFi+LibriTTS - additionally adding LibriTTS corpus, which results in approximately 155000 utterances in the training set.
We evaluate both scenarios using a MUSHRA subjective naturalness test analogously to Section \ref{ssec:experiments_text_to_speech_naturalness} and summarize results in Table \ref{table:mushra_naturalness_hifi_predicted}.
Using an additional, large-scale dataset during training significantly improves naturalness of synthesized speech ($p$-$value<0.01$).
Per-case analysis of listeners judgements reveal that both systems produce audio of similar segmental quality, but the HiFi+LibriTTS system provides more stable prosody, especially for longer utterances.
We conclude that semantically concerted prosody prediction is a data hungry problem and limited amount of training data can result in less stable prosody of generated speech.
However, using auxiliary data in the training procedure allows to obtain a more robust prosody predictor and mitigate this issue. 

\begin{table}[htb]
\centering
\begin{tabular}{lll}
\toprule
\textbf{HiFi} & \textbf{HiFi+LibriTTS}    & \textbf{Ref}   \\ \hline
\(64.26\)           &\(67.01\)                        & \(67.17\)      \\ 
\bottomrule
\end{tabular}
\caption{\textit{Mean MUSHRA scores for the word-level models trained on a single speaker form the HiFi corpus with and without auxiliary LibriTTS data. All comparisons from this Table are statistically significant ($p$-$value<0.01$).}}
\label{table:mushra_naturalness_hifi_predicted}
\vspace*{-3mm}                                     
\end{table}

\subsection{Prosody Predictor Input Streams Ablation Study}
\label{ssec:experiments_prosody_predictor_input_streams_study}
\begin{figure}[htb]
\begin{minipage}[b]{1.0\linewidth}
  \centering
  \centerline{\includegraphics[width=8.5cm]{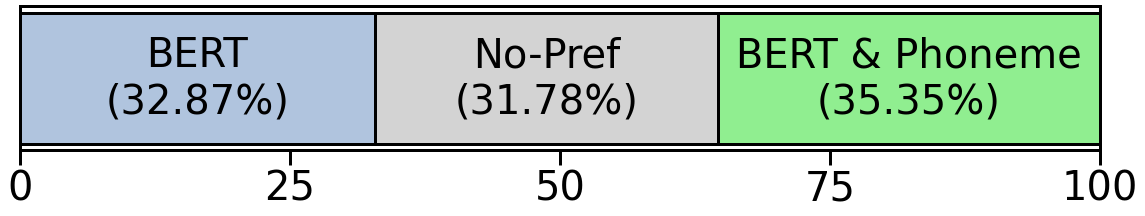}}
  \centerline{\textit{(a)}}\smallskip

\end{minipage}
\begin{minipage}[b]{1.0\linewidth}
  \centering
  \centerline{\includegraphics[width=8.5cm]{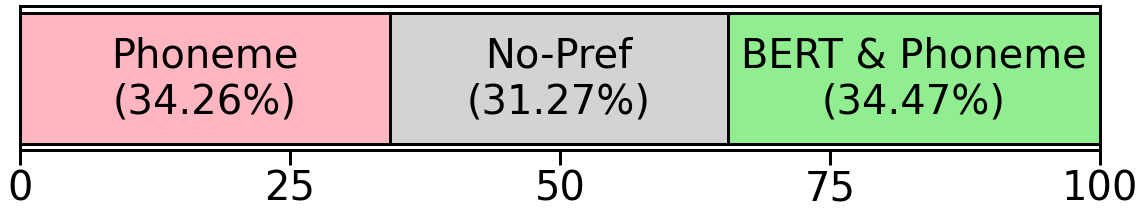}}
  \centerline{\textit{(b)}}\smallskip
\end{minipage}

\vspace*{-3mm}                                     
\caption {\textit{Results of naturalness preference tests. On the right the word-level prosody embeddings predictor model described in Section \ref{sec:prosody_embeddings_predictor}. On the left the same model but with only one input stream: a) BERT embeddings, b) phoneme sequence.}}
\label{fig:pref_prosody_predictor_ablation}
\vspace{-3mm}
\end{figure}

In previous works, word-level prosody embeddings are typically predicted at inference time from one of the following representations derived from text: word-level contextual embeddings \cite{hodari2021camp, ren2022prosospeech} or phoneme sequence \cite{klapsas2021word, guo2022unsupervised}.
We use both of them in our prosody embeddings predictor model.
To determine the contribution of each input stream to the model performance we conduct an ablation study.
We train the word-level prosody embeddings predictor model in three configurations: \textit{BERT \& Phoneme} - trained exactly as described in section \ref{sec:prosody_embeddings_predictor}; \textit{BERT} - with only BERT embeddings input; \textit{Phoneme} - with only phoneme sequence input.
All three predictor models are used in combination with the same W-VAE acoustic model to synthesize 1000 validation utterances from section \ref{ssec:experiments_data}. 
As a subjective naturalness evaluation, a preference test is carried out using Amazon Mechanical Turk platform. 
Native English speakers are asked to \textit{"Select which audio sounds more natural"} for pairs of audio samples generated with different systems.
We find that removing fine-grained phoneme sequence input stream (Figure \ref{fig:pref_prosody_predictor_ablation}a) causes less stable prosody prediction and results in significant degradation in naturalness ($p$-$value<0.01$). 
Whereas, the difference between BERT \& Phoneme and Phoneme systems (Figure \ref{fig:pref_prosody_predictor_ablation}b) is not statistically significant.
Listening to samples revealed that the improvement of using contextual word embeddings comes mainly in better phrasing and pause prediction.
However, differences are quite subtle and therefore are not reflected in the crowdsourced naturalness evaluation results.
This result contradicts the findings from \cite{hodari2021camp}, where significant quality degradation when removing BERT embeddings input was reported.
However, in \cite{hodari2021camp} authors proposed to use word-level syntax features e.g. part-of-speech labels,
compound-noun flag and punctuation flag as a second input stream to the model predicting prosody embeddings.
We believe that fine-grained phoneme sequence input is more correlated
with prosody than those syntax features, therefore, removing BERT input stream is less harmful in our work.

\vspace*{-2mm}                                     
\section{Conclusions}
\label{sec:conclusions}
In this paper we explored the design of prosodic representations learned in an auto-encoder manner.
First, we introduced a TTS framework allowing for a fair comparison of prosody embeddings of different granularity.
Then a systematic study of utterance, word and phoneme-level representations was conducted on a large-scale, publicly available dataset - LibriTTS.
Through our experiments, we demonstrated the trade-off between capacity and predictability of prosody representations.
We showed that utterance-level embeddings have insufficient capacity to model expressive speech variability.
Whereas phoneme-level representations require strong regularization for stable prediction from text at inference time.
We found that word-level embeddings impose a good balance between capacity and predictability.
As a result, we closed the gap in naturalness by 90\% between synthetic speech and recordings  without sacrificing intelligibility.
Finally, we looked into applying the proposed approach in a single-speaker scenario.
We showed that semantically concerted prosody prediction is a data hungry problem and limited amount of training data can result in less stable prosody of generated speech.
However, this issue can be mitigated by using auxiliary data in the training procedure.

\bibliographystyle{IEEEbib}
\bibliography{refs}

\end{document}